\documentclass[orivec]{llncs}

\usepackage{graphicx,url,amsmath}

\newcommand{\hdb}{\emph{hdb} }
\newcommand{\hdbt}{\emph{hdb}}

\newcommand{\sw}[1]{\emph{#1} }
\newcommand{\swt}[1]{\emph{#1}}

\newcommand{\pl}[1]{{\url{#1}}}

\begin{document}
	\title{An extensible web interface for databases and its application to storing biochemical data}
	\author{Nicos Angelopoulos \and Paul Taylor}
	\institute{
		Department of Biological Sciences, University of Edinburgh, Edinburgh, Scotland, UK \\
		\email{\{n.angelopoulos,p.taylor\}@ed.ac.uk}
	}
\maketitle

\begin{abstract}
   
   This paper presents a generic  web-based database  interface implemented
   in Prolog.  We discuss the advantages of the implementation platform and
   demonstrate the system's applicability in providing access to integrated
   biochemical  data. Our system  exploits two  libraries of  \sw{SWI-Prolog} to 
   create a schema-transparent interface within a relational setting. As is
   expected  in declarative  programming,  the interface  was written  with
   minimal programming effort due to the high level of the language and its
   suitability to the task. We highlight  two of Prolog's features that are
   well  suited to  the task  at  hand: term  representation of  structured
   documents and  relational nature of Prolog which facilitates transparent
   integration  of relational databases.  Although we developed the  system
   for  accessing in-house  biochemical and  genomic data the  interface is
   generic  and provides a number  of extensible features. We describe some
   of these features with references to our  research databases. Finally we 
   outline  an  in-house  library  that facilitates interaction between 
   Prolog and the \sw{R} statistical package.  We describe how it has been 
   employed in the present
   context to store output from statistical analysis on to the database.


\end{abstract}

\section{Introduction}

Declarative programming in general and logic programming (LP) in particular
when compared to other paradigms, present a much higher level at
which programs can be composed. The resulting programs are typically
written with less programming effort and are easier to understand.

Interfaces for relational databases are usually the subject of
corporate development and absorb substantial programming effort.
It is often the case, that the languages these are implemented in
are more suitable for the graphical aspects of presentation
rather than for capturing the underlying declarative model of the databases.
The HTML language, (see for instance \cite{Html99}),
presents a structured approach to user interaction. By design,
it is meant for the focus to be on the content and the relations 
across content rather than on exact graphical coordination.

We constructed a high level web-based interface by exploiting
two libraries of the \sw{SWI-Prolog} engine. 
The first library facilitates the seamless integration of 
databases via the \sw{odbc} interface. Prolog is particularly well 
suited both for manipulating database meta-data (table structure
and fields) and for reasoning with the primary data.
The second library presents a uniform way for translating between terms
and HTML code and an interface for low level communication with the
operating system.
The former facilitates web-page composition via term manipulation
while the latter deals with non-logical aspects of the interaction with
the web-services. In tandem the two libraries allow a comfortable
high-level style of programming used for rapid-prototyping and 
the refinement to a stable system that is fully implemented in Prolog.

Logic programming for web services have been advocated in a number
of approaches. Notably, in the Pillow library 
(\cite{CabezaD_HermenegildoM_2001}) which is available 
for a number of logic engines. Representing HTML code as term
structures is an appealing proposition. Prolog programs can 
construct these at run-time with correct HTML code generated
by library predicates.

The suitability of HTML as an interface platform for databases is not 
unique to our approach. It is a very popular choice particularly 
through the PHP scripting language, \cite{Php5}. The benefits of our approach 
relative to PHP is its relational and high-level nature and 
the fact that Prolog is a general purpose programming language.
On the other hand PHP provides more dedicated features and 
a large community of practitioners.

The paper is structured as follows. The main components of the
system are presented in Section 2 followed by its main
database-independent functions in Section 3.
Section 4 presents some specific characteristics of using
the interface for biochemical data.
The concluding remarks are in Section 5.

\section{System architecture}

\begin{figure}[t]
  \includegraphics[width=1\textwidth]{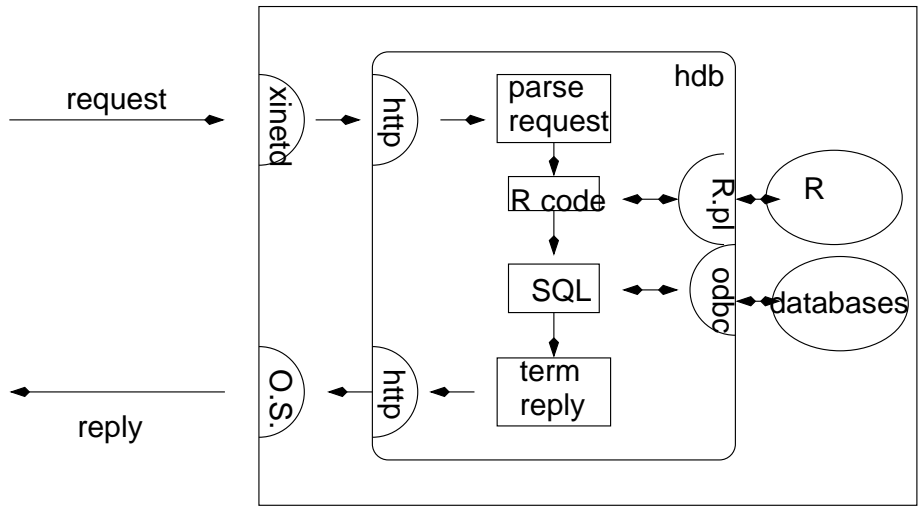}
  \caption{System architecture and interactions} 
  \label{fig:arch}
\end{figure}
   
This section describes the architecture of our system (\hdbt) and
its interaction with the operating system. 
The overall flow of data is shown in Fig.~\ref{fig:arch}. 
Semi-circles show interaction software between systems.
The round-edged rectangular labelled `hdb' depicts our system
and the three rectangular boxes show its main components.

Requests to service a page arrive at the server machine.
The operating system invokes \sw{hdb} via \sw{xinetd} and the \sw{http}
library is used to collect the request. This is parsed and 
in turn \sw{SQL} or queries about meta-data are constructed.
Using \sw{odbc} the system interacts with the databases and
the results of this interaction are processed into term structures
that will form the HTML page to serve. The page is generated
by the \sw{http} library. In the specific application we have
employed the generic interface, \sw{http} requests are also parsed 
for references to statistical analysis to be performed by \swt{R}
\cite{R2006}.
This is typically the scenario when either a new mass spectrometry 
experiment is uploaded, or a comparative analysis table entry is 
requested by the user.
The \sw{R.pl} library will handle any such interactions communicating
any results from the analysis that need to be saved, to the SQL 
generator.

The system presented here is fully implemented within \sw{SWI-Prolog}
(see \cite{WielemakerJ_2007}).
The two crucial parts of this Prolog engine that enable this approach
are the \sw{http} and \sw{odbc} libraries. As can be seen in 
Fig.~\ref{fig:arch} the \sw{http} library plays two complementary
roles. On one hand it provides the server machinery for reading 
in requests, while its other role is to translate special Herbrand terms
to HTML documents. The full capabilities of \sw{SWI-Prolog} with regard
to web-services is discussed in detail in \cite{WielemakerJ+2007}.

The \hdb server is based on the \sw{http} library and can be started
in two ways.
The simpler method is by presenting a call to the engine, such as:
\[ \mbox{\pl{ ?- http_server(reply, port(8080),timeout(30)] ).}} \]
As long as the engine that runs this query is active, the port 8080
will be serviced by the predicate \pl{reply(+Request)}.
The library instantiates \pl{Request} to a Prolog representation of the 
incoming request.
This method is particularly useful during software development as
it provides a terminal at which messages about the computation in
progress can be delivered to.

\begin{figure}[t]
\begin{tabular}{llcl}
service hdb \\
\{ \\
        & port            & = & 8080 \\
        & socket\_type     & = & stream \\
        & protocol        & = & tcp \\
        & wait            & = & no \\
        & user            & = & nicos \\
        & server          & = & /srv/www/html/hdb/hdb\_xinetd \\
        & log\_on\_failure  & += & USERID \\
        & log\_on\_success  & += & PID HOST EXIT \\
\} \\
\end{tabular}
\caption{A typical \sw{hdb xinetd} file}
\label{fig:xinetd}
\end{figure}

Alternatively and more conveniently for non-developmental deployments,
the server can be started through
an intermediary piece of software such as \sw{inetd} or as is common
in our set-up and as illustrated in Fig.\ref{fig:arch}, by \swt{xinetd}.
These are daemons in the terminology of operating systems, programs 
that run continuously listening to the internet ports.
A typical \sw{xinetd} entry will reside in \url{/etc/xinetd.d/hdb} and
contain the entries shown in Fig.~\ref{fig:xinetd}.

Requests to the relevant port (8080 in our example) need to be 
allowed connection to the server. Typically this is via an entry to the 
services (\url{/etc/services'}) file such as 
`hdb         8080/tcp' \cite{PostelJ_80}.

Once the request has been passed to \hdb and parsed, generation of a 
reply page will 
typically instigate some interaction with the serviced databases. 
The actual interaction is facilitated by the \sw{odbc} library.
There are two major types of requests to be served. 
Ones that require meta-data and ones that manipulate primary data.

The facilities of the \sw{odbc} with regard to meta-data are 
particularly useful as it means the core interface is completely
free of references to any specific database.
For instance by using:\\
\begin{eqnarray*}
  \lefteqn{} & \lefteqn{} & \mbox{\pl{odbc_current_connection(-Connection, -DSN)}}
  \\
  \lefteqn{} & \lefteqn{} & \mbox{\pl{odbc_get_connection(+Connection, database_name(-DB))}}
\end{eqnarray*} \\
all open connections can be found and via that all the related databases.
Similarly, use of \\
\begin{eqnarray*}
  \lefteqn{} & \lefteqn{} & \mbox{\pl{odbc_current_table(+Connection, -Table)}}
  \\
  \lefteqn{} & \lefteqn{} & \mbox{\pl{odbc_table_column(+Connection, +Table, -Column)}}
\end{eqnarray*} \\
provides access to the database dictionary. With these calls,
the database structure and standard operations can be displayed
without any hard-wired dependencies. In the following Section we will 
show in detail how these are used in our system.
The \sw{odbc} library depends on operating system connectivity to \sw{ODBC}
(open database connectivity) software. An instance of such software
is \sw{unixODBC} \cite{unixODBC2007}. In our experience this has worked well
with the Prolog libraries discussed here.

The organisation of the code reflects the two main components of the 
system. Core functionality predicates are stored in directory \pl{src}
while extensibility predicates are in directory \pl{site}. 
The distributed code is ready
to be deployed as soon as a minimal site specific configuration is set-up. 
This should describe the location of the databases to be served and 
the users of the interface. Note that no information about the 
database schemata is needed for the core interface to operate
correctly.

Our final note in this section is that \sw{SWI-Prolog} in tandem with other
open source software provides a powerful platform for serving HTML pages. The
high-level of programming, the independence of specific data-sources and the
structured nature of Herbrand terms which maps well to correct HTML code 
ensure that effective interfaces can be built with minimal programming effort.

\section{Core functionality}

In this Section we describe the main functions of the interface.
We present and discuss features that are generic and can be employed 
as-is in any installation of the system. We distinguish six types of pages 
in this category: (a) authentication, (b) overview of databases
(c) view of database, (d) single table, (e) single view and (f)
profile data. In what follows we detail these six categories.

The first point of contact a user will have with the interface
is the authentication page, where they are asked for a username
and password. 
There are two types of authentication services provided.
The site administrator must choose one of them at installation.
The first, uses the Prolog internal mechanism that keeps a thread
alive for a specified amount of time. Usually the amount of time is set to
a few minutes. Within that period
the connection is authenticated and the user can interact with
the system. Each interaction resets the thread to the start of the
interval. If there is inactivity for a period longer than the interval
the process dies and the user will need to log-in again.

\begin{figure}[t]
  \includegraphics[width=1\textwidth]{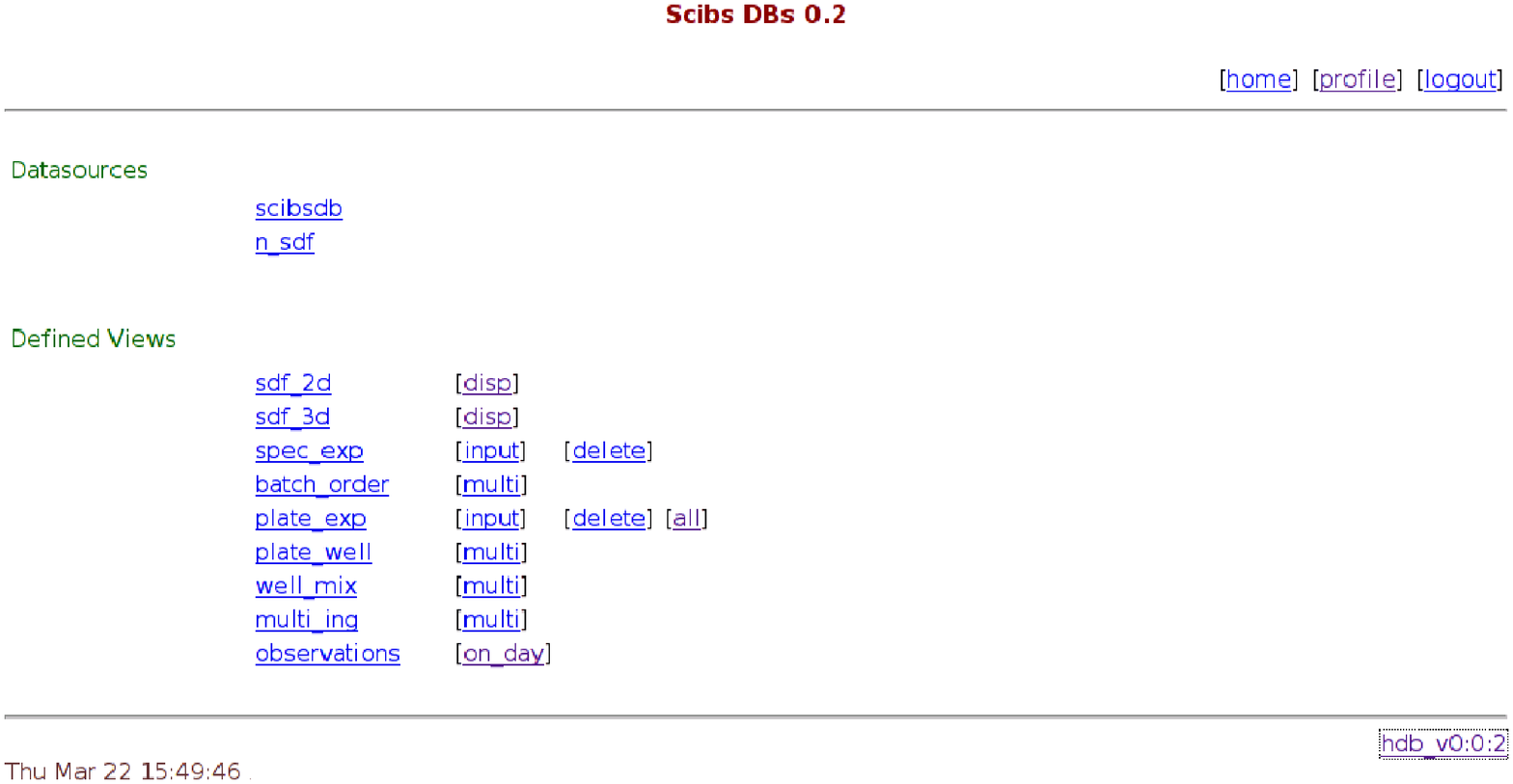}
  \caption{Home page} 
  \label{fig:home}
\end{figure}

The second type of authentication services is IP address based. 
In this mode of operation, the credentials of a user are associated to 
the IP address from which they last log-in. The interval during which this
activation is valid is usually set to a few hours, typically 
the duration of a working day. This provides a less
secure model but which is much easier to use in practice.
In both authentication scenarios, \hdb is by necessity of a more or equal 
restrictiveness to that of the underlying databases with regard to 
database operations.
Each \hdb user is assigned to a db user in a modular way
and the exact mapping is a matter of site administration, although
typically both credentials will be identical.
It is worth noting that authentication has been designed with ease of
use in mind rather than maximal security from determined attacks.

Once users have tackled authentication and gained access to the system they
are presented with the home \hdb page which shows an overview
of the resources they can access. An example is shown in Fig.~\ref{fig:home}.
At the top right, there are persistent navigation options that allow 
logging-out, access to session profile and a link to the home page.
The main body of the page gives access to the databases and views 
accessible to the user. 

Each database page presents the tables which are contained in the 
database source along with a subset of standard operations that can
be performed on each. A partial example is shown below:

\begin{tabular}{llllll}
Experiment		& [input]& 	 [update]	&  [delete]& 	 [query]& 	 [all] \\
ExternalDataSource& 		[input]& 	 [update]& 	 [delete]& 	 [query]	&  [all] \\
Input& 		[query]& 	 [all] \\
Mix	& 	[input]& 	 [update]	&  [delete]	&  [query]	&  [all] \\
MixIngredient& 		[input]& 	 [update]& 	 [delete]& 	 [query]	&  [all] \\
Plate	& 	[input]& 	 [update]	&  [delete]& 	 [query]& 	 [all] \\
\end{tabular}

\noindent Note that in our example table `Input' is read-only and it is
thus only associated with read operations. This table
is dedicated to recording the changes that occur to the databases from
within the \hdb interface.
Our system supports explicit storage of such information. 
Navigating via the table name will present a single table page while
the link on each operation links to the operation page for the 
specified table.

A page detailing a single table shows the fields of the table
along with their types and provides access to operations on the table.
These are identical to those that appear along side the table name in the 
database page. They will usually be a subset of: `input', `update', 
'delete', 'query' and 'all'. Typically it will be the maximal subset,
and it is also possible to hook table or database specific operations.
An extract from a table page is as follows:

\begin{figure}[t]
  \includegraphics[width=1\textwidth]{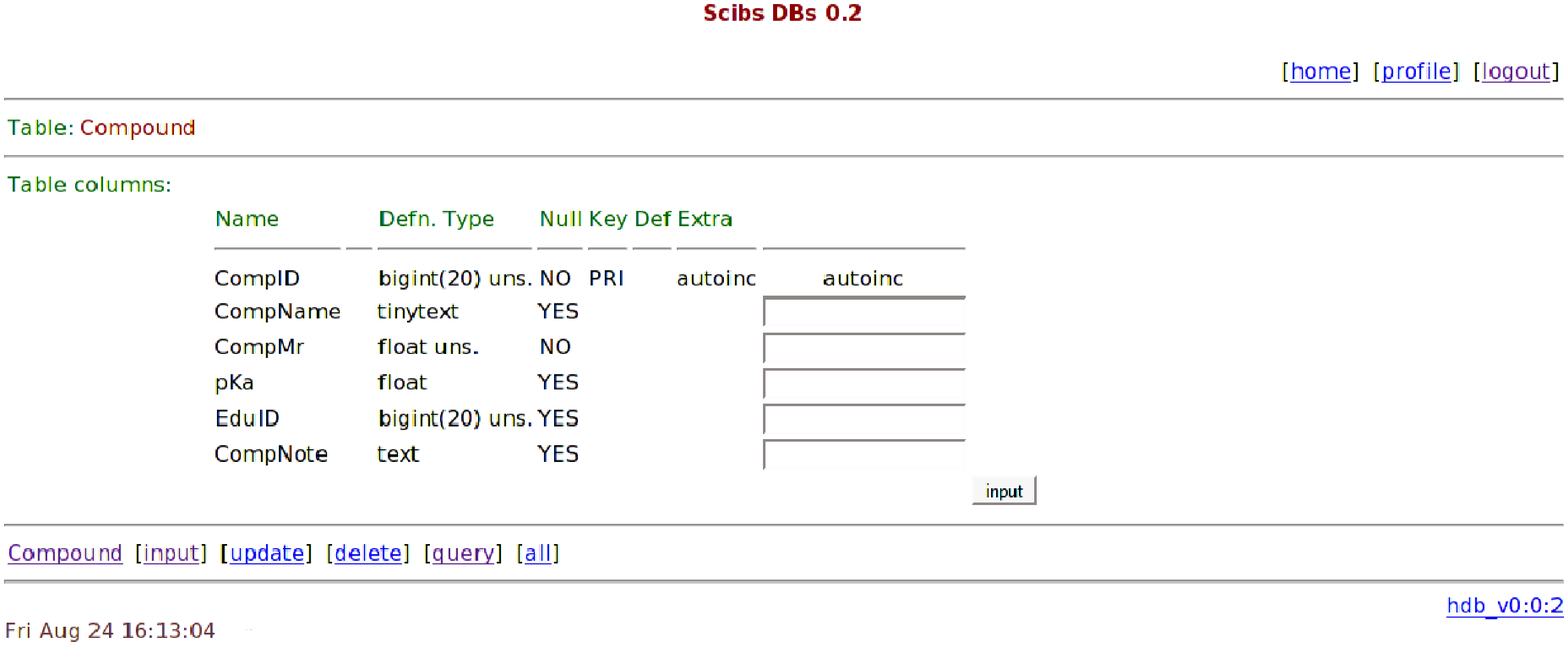}
  \caption{Input data to a table.} 
  \label{fig:input}
\end{figure}

\vspace{0.3cm}
\begin{tabular}{llllll}
scibsdb.Compound	 & has 210 rows. \\
Table columns:	 \\
Name	      &	Defn. Type     & Null	& Key  & Def & Extra \\
CompID      & bigint(20) uns.	& NO	   & PRI	 &	    & autoinc \\
CompName		& tinytext	      & YES	\\
CompMr		& float uns.	   & NO  \\
pKa		   & float	         & YES	\\
EduID		   & bigint(20) uns.	& YES	\\
CompNote		& text	         & YES \\
\end{tabular}

\vspace{0.3cm}
The first line gives the database and table name along with its 
population size.
The headings of the columns appearing on the third line of the example
above are, in left-to-right order: the field name, the type of the field,
a Boolean value signifying whether or not a null value is allowed, 
 a value reflecting whether field is a key for the table,
 the default value for the field and finally, any extra information about
 the field.
In our example the first field, `CompID', is an auto-incrementing 
integer field. Also note that `uns' is an abbreviation for unsigned.

Most operations follow intuitively after their name. The `all' operation
is the maximal query that allows the user to view all data in the table.
(Standard site-wide defaults of presentation style apply.) One of the more
useful operations is inserting data into tables. An example is shown 
in Fig.~\ref{fig:input}. The interface automatically handles auto increment
fields (integer fields which are routinely used as unique identifiers) and 
enumeration types that are displayed as dropped down menus.

\begin{figure}[t]
  \centerline{\includegraphics[width=0.8\textwidth]{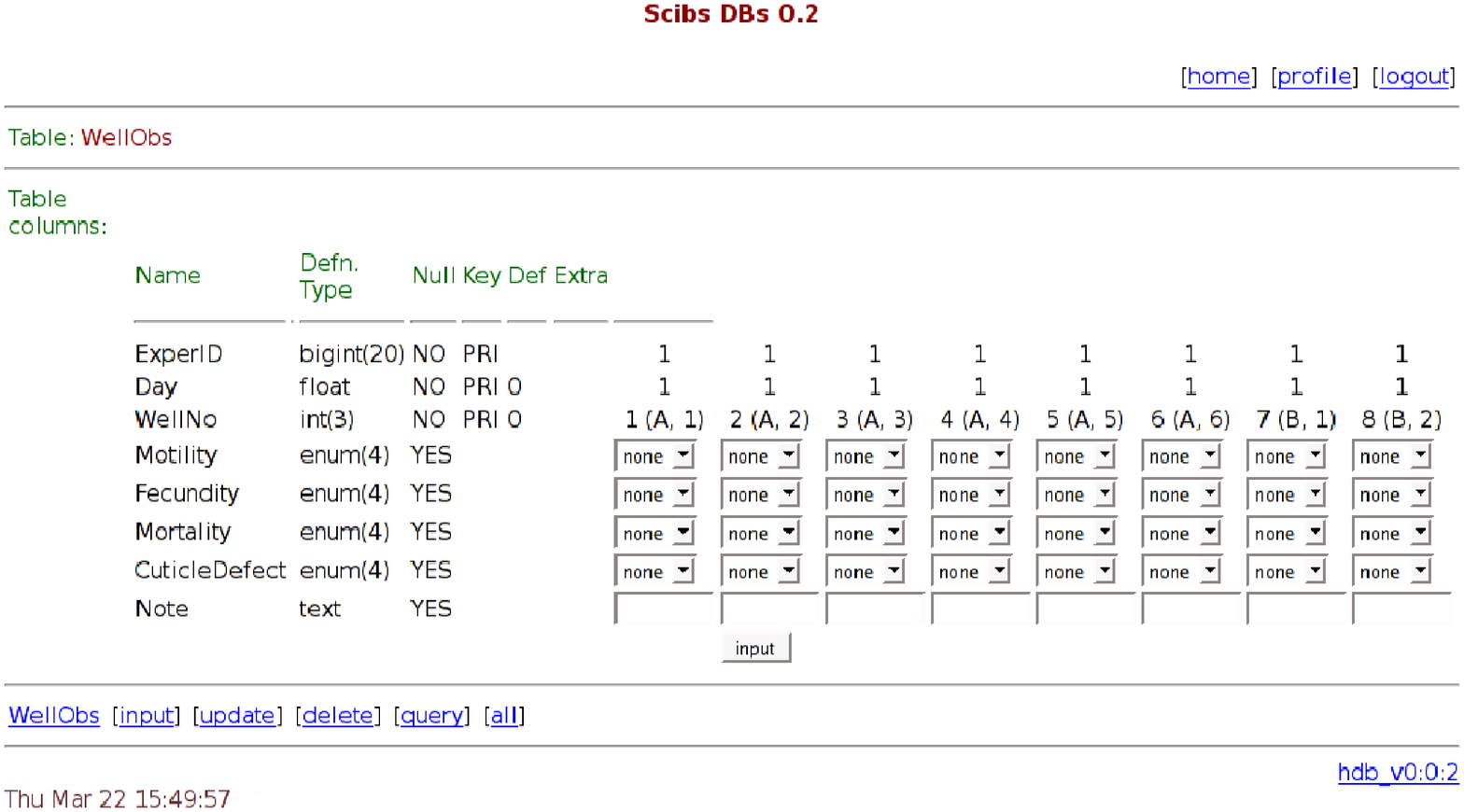}}
  \caption{Input of multiple experiments.} 
  \label{fig:obs}
\end{figure}

Returning to describing the `home' page, we have another type
of object, namely, views. Views are convenient higher level 
objects which allow operations that 
manipulate more than one table, or more than one row of
a single table within a single step. In Fig.~\ref{fig:home} 
a number of specialised views are offered to the user.
For instance `sdf\_3d' displays the 3D representation of an SDF entry. 
Each SDF is a textual representation of a chemical substance 
(\cite{DalbyA+1992}).
The view `observations' has one operation which allows
results from many experiments taken at a single time-point
to be inputted together. As most of the views and operations are 
specific to a given site and makes extensive use of the extensibility
features of our system they will be discussed in further detail
in the next section. An example of an operation to such
a view is given in Fig.~\ref{fig:obs}, the details of which will also 
be discussed in what follows.
It suffices here to say that \hdb provides convenient hooks for new
views and associated operations to be defined within Prolog.

Finally, the user has access to information profiling his session. 
Information regarding the user names, IP address, session ID and 
the server associated with this session is presented, as illustrated
in the following example:

\begin{tabular}{l}
\\
User Profile \\
 \\
Logged-in on hdb server: Scibs DBs 0.2 \\
With user name: nicos \\
Database user name: nicos \\
Login time: at(2007, 8, 24, 14, 22, 40, 10) \\
Peer: 129.215.137.168 \\
 \\
Pages are served by : hdb  0:0:2 \\
Server: scibsfs.bch.ed.ac.uk:8080 \\
Session: 5807-da08-fbaa-fe69 \\
\end{tabular}
\vspace{0.5cm}

Warnings and errors about \hdbt's operation are reported by the
system at the top of the first generated page after the error was 
caught. Messages are asserted as \pl{diagnostic/1} terms via the
\pl{http_session_assert/1} primitive. The example message 
\[ \mbox{\pl{unable_to_connect_to_db_source(nilhhloc-ni_lhh)}} \]
informs of the unavailability of database \pl{ni_lhh} 
right after log-in. The message appears near the top of the page
in orange colour as to draw the user's attention and is non persistent.

\section{Storing biochemical data}

The interface can be tailored to a site's specific needs through a
number of hooks and handlers. Both of these can alter the HTML generated
by the system.
Hooks are optional parts that when present alter a specific part of the
interaction. For instance, a particular field's output can 
be linked automatically to a live HTML link, or certain fields for
a table are auto-filled by some scripts. Hooks are appropriate
in extending the system in a way that allows database table evolution.
The addition of extra tables or columns (fields) will usually require
no changes to hooks of existing columns.
Handlers on the other
hand, are predicates that deliver more substantial extensions to 
the interface.
For instance a handler may produce the HTML code from a non-standard
operation on a view (such as the `disp' operation on view `sdf\_3d'
detailed below).

\begin{figure}[t]
\begin{verbatim}
% hdb_hook_column_input_def_value(+DB, +Table, +Column, -Def ).
hdb_hook_column_input_def_value(ni_lhh, _, Column, Date) :-
     atom_concat( _, 'Date', Column),
     get_time(Time),
     convert_time(Time, Yr, Mo, Dy, _Hr, _Mn, _Se, _Ml),
     number_codes(Yr, YrCs),
     number_codes(Mo, MoCs),
     number_codes(Dy, DyCs),
     flatten([YrCs,"-",MoCs,"-",DyCs], DateCs),
     atom_codes(Date, DateCs).
\end{verbatim}
\caption{Date default values in database `ni\_lhh'.}
\label{fig:defvalue}
\end{figure}

For example, hooks are used in views to declare the participating columns 
(\pl{db_view_has_column( +View, +DB, +Table, +Columns, +JnKeys)})
and the operations defined on a particular view 
(\pl{db_view_has_ops( +View, +Op )}). For tables, hooks are very useful 
in defining input, output settings that can take into account 
local administration issues. For instance, 
\pl{hdb_hook_column_input_textarea(+DB, +Table, +Clmn, -Rows, -Cols)}
can be used to overwrite the default size for the input text box at
an insert operations.
For a more concrete example, consider the code in Fig.~\ref{fig:defvalue}.
Predicate \pl{hdb_hook_column_input_def_value(+DB, +Table, +Column, -Def )}
when defined for a specific column, table and database combination, it
dictates a default input value for the said column. The shown code
defines the current date as the default value for each column in database
`ni\_lhh' that has a name containing `Date' as a suffix. In our example this
is true irrespective of the table's name.

In the remainder of this Section we will present some database-specific
aspects of the system and our local installation which
stores lab results and biochemical data.
Briefly described the main requirements are:
storage of experimental data from (a) C. elegans growth, 
(b)micro-array and spectrometry data for samples resulting from the 
C. elegans experiments, and (c) information about associated molecules.

\begin{figure}[t]
  \centerline{\includegraphics[width=0.6\textwidth]{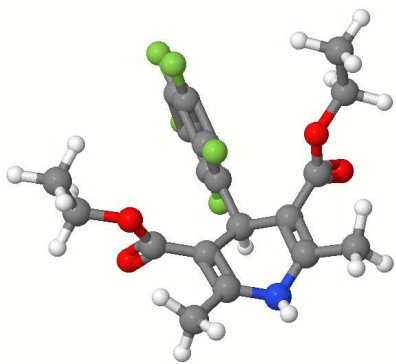}}
  \caption{Jmol invocation. NemadipineA, (hdb) Eduliss ID: 108525. } 
  \label{fig:jmol}
\end{figure}

Although a limited number of target ligands are used in our experiments
we are able to include into our system a much broader
selection of chemical substances. In total 1.7 million are available 
and they are drawn from the EDULISS database \cite{HintonCA_2005} (version 1.0).
A structured schema has been devised for
storing the atomic information that reside in flat format SDF files. 
Individual molecules can then be inspected by retrieving all relevant
entries writing those out in the usual flat file format and call 
the Jmol java applet, \cite{Jmol}, to display the three or two dimensional representation 
of the molecule. An example of a displayed ligand (small molecule) is in 
Fig.~\ref{fig:jmol}. The display operation is labelled `disp' and is
available for views `sdf\_2d' and `sdf\_3d'. The example shown is 
Nemadipine-A and was shown to induce a marked phenotype in C. elegans
(\cite{KwokTCYK+2006}).

It is often the case that complications of database interfaces 
arise from the disparity between the best way to hold the data
in the database and the users' way of organising the data in their minds.
Inputting data is one crucial part in which a reconciliation is vital.
In our system views can be used to insert data in ways that are more
intuitive and faster for the user. 
As an example consider Fig.~\ref{fig:obs}.
The user is able to input a number of observations corresponding 
to `well' experiments that were started at the same time. This is an
intuitive way of viewing the set of experiments as all wells
come from a single plate and have synchronised start times.
On submitting the
data a number of separate table entries will be stored in the database. 
The majority of the entries in Fig.\ref{fig:obs} are in the form of
selection menus as they are of enumeration type further easing
the input process.

A major consideration in experimental settings is the management of 
raw datafiles as well as the tracking of large files as they 
go through various stages of analysis.
Our approach is to use the uploading capabilities of the \sw{http} library
and store the files in canonical locations with a file link stored
in the actual database. The library has been performing robustly and
has coped with files of substantial sizes. 

\begin{figure}[t]
   \begin{tabular}{cc}
   \includegraphics[width=0.6\textwidth]{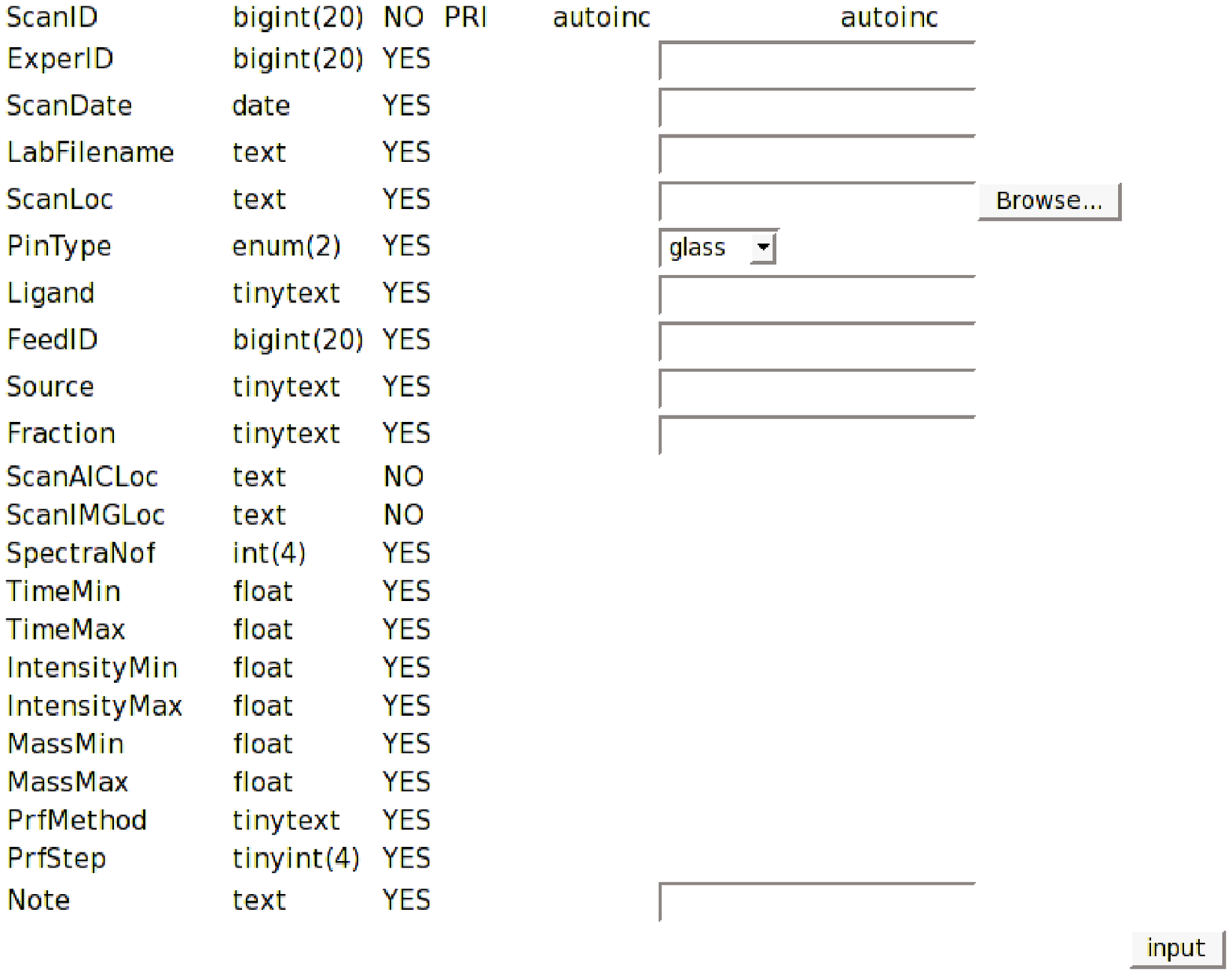} & 
   \begin{minipage}[b]{0.35\textwidth}
   \includegraphics[width=1\textwidth]{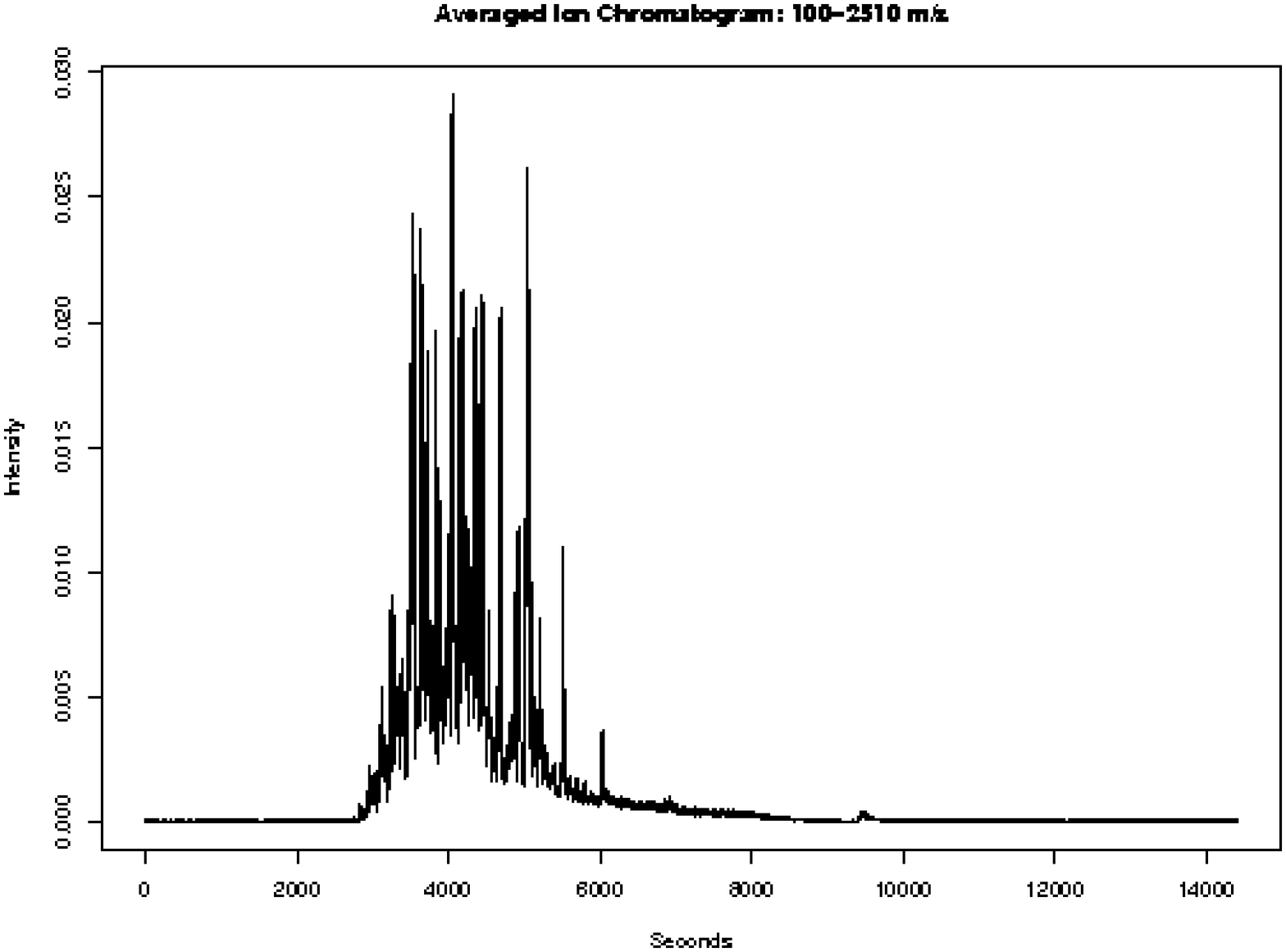}
   \\
   \includegraphics[width=1\textwidth]{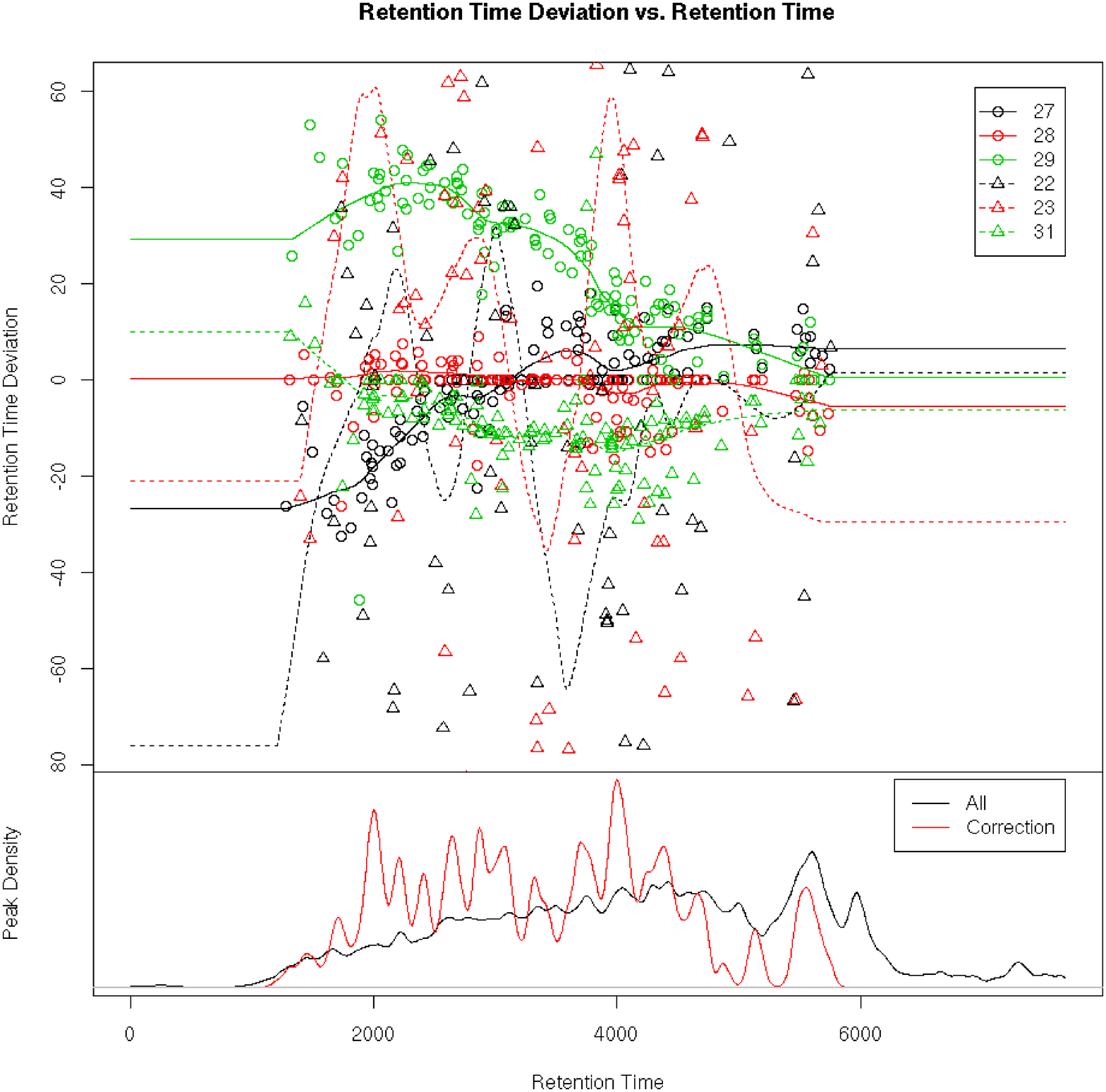}
   \end{minipage}
   \\
   \end{tabular}
   \caption{Left: automatically filled fields. Right: generated averaged ion count (top) and retention time correction (bottom).}
   \label{fig:autofill}
\end{figure}

In addition to storing primary data we have used auto filling capabilities
in the system to automatically generate store derived information on the 
data. In particular output from mass-spectrometry experiments that are 
exported in NetCDF format are uploaded to the database. The 
\sw{xmcs} \cite{xcms} part of the \sw{Bioconductor} \cite{Bioconductor2007} 
package for the \sw{R} \cite{R2006} statistical system is started 
through a Prolog interface. We have built a simple \sw{Prolog} to $R$ 
interface that runs the latter as a slave with the command:
\begin{tabular}{@{\hspace{1cm}}l}
\\
\verb+R --slave --no-environment+
\\
\vspace{0.02cm}
\end{tabular}

\noindent The input, output and error streams of this process are handled by 
Prolog as to enable interaction with the \sw{R} shell. The in-house library
facilitates the translation of term structures to atoms that are 
written on to the \sw{R} process' input. Also it reads in and translates
output that appears in the \sw{R} process' output and error streams.
It thus allows Prolog to access the vast wealth of functions
and packages available in \swt{R}. The Prolog-to-R interface is also made
available as an independent module (\cite{rsession})
and is included into \sw{SWI} main source distribution
as a contributed package.

Fig.~\ref{fig:autofill} shows an example of automatically filled table entries.
In entering data to table `SpecScan' the user can only input values for 
7 out of the 18 fields. The remaining are filled with: 
(a) visual summaries of the data- a postscript output of the averaged
ion count for each scan (ScanAICLoc) and a heatmap image (ScanIMGLoc),
(b) information about the sample that exist in the bundled input
(fields SpectraNof to MassMax) and basic parameters for some of the 
visual summary fields- here PrfMethod	and PrfStep for the heatmap generation.
All derived information are produced by invoking the \sw{xcms} \sw{R} package
on the data uploaded by the mass-spectrometrist (field ScanLoc).

Our approach has the benefit that output from specialised software
is stored in accessible forms that can be viewed by all partners
in the project.
It is also the case that it improves quality control tasks 
and that by automatically filling the form
there is less mundane typing and data-entry and thus less chances of 
an error to occur. Additionally in the scenario we have employed the 
interface, the statistical analysis with \sw{xcms} takes a substantial 
amount of time to run. By storing the results of the analysis in a central 
repository with all associated information managed automatically, we 
ensured that subsequent inspections of the analysis are handled
promptly and correctly.

The top left of Fig.~\ref{fig:autofill} shows the main part of an input 
operation with the values of the fields that can be changed filled.
At the top right is the generated averaged ion count (AIC) which 
plots average ion count against time and provides an overall picture
of activity within the spectra. The bottom right of Fig.~\ref{fig:autofill}
shows the retention time correction as generated by \sw{xcms}
when creating a comparison among spectra from two conditions.
Retention time correction is performed as to align multiple 
mass-spectra from two distinct conditions so that their peaks can 
be compared in a meaningful way. Visual inspection of the correction 
is a good indicator for the quality of the alignment and thus a 
crucial information to include in the stored derived information information. 



   
\section{Conclusions}

We presented a system that implements a high level interface for 
data sources by exploiting two \sw{SWI-Prolog} libraries. The two main 
advantages of our implementation platform are the relational
nature of Prolog and the compositional nature of Herbard terms.
The former allows seamless integration of relational databases
and the latter facilitates construction of structured HTML
code.

To our experience the main benefits of \sw{hdb} are (a) that it implements
a well separated organisation of what is generic and what must be
tailored for a particular installation, and (b) that the generic part 
of the system is schema-driven in the sense that it is automatically
constructed from the underlying databases with no hard-wired dependencies.

Our work has demonstrated the usefulness of a relational
programming language for providing interfaces to databases via served
HTML pages. The generic part of the interface can be deployed directly
to any existing database. The installation can then slowly be ramified
and tailored to the particular site via a number of extensibility
features the system provides. 

We have identified four directions in which we will direct 
future work. First is the ability to have communication between different
pages via collections of database items. These can be through tick boxes
at query result pages. Also to provide generic mechanisms for 
background processes filling of derived information. This will allow 
the interface to work as a front-end and hiding a lot of dedicated
background processing. In our databases, for instance, we can allow
mass-spectrometry staff to create views that explore differentially
expressed mass over charge peaks from a set of controls against a set
of screens. The third direction in which to develop the system is 
in providing simpler query mechanisms such as ability to choose a 
list of fields and provide values or regular expressions for those.
Finally, to modularising the database access and
allowing alternative methods of interaction such as via the BerkleyDB
interface (package `db' in \swt{SWI-Prolog}).

The source for the system described in this paper can be downloaded from
\url{http://scibsfs.bch.ed.ac.uk/~nicos/sware/hdb}. It has been developed
and tested on Linux systems.
The in-house library 
that handles communication with \swt{R}, \sw{r\_session,}
is included as a contributed package in current releases of
\swt{SWI-Prolog} (\swt{library('R')}).

\section{Acknowledgements}
This work was funded by BBSRC via grant BB/D00604 X/1 which is part of
the SCIBS (Selective Chemical Intervention in Biological Systems) 
initiative.
We would like to thank Prof Malcolm Walkinshaw for his encouragement 
and project leadership and Kun-Yi Hsin for help with the SDF files 
and EDULISS database. We are indebted to Jan Wielemaker, the developer
of \swt{SWI-Prolog},
for developing a very stable and powerful core system with two exemplary
libraries. 
Also we thank him for his prompt replies in resolving technical issues
and for included our \sw{R} interface to the \swt{SWI}'s main source branch.

\bibliography{submit}
\bibliographystyle{apalike}

\end{document}